\documentclass[twocolumn,nofootinbib]{revtex4-1}
\usepackage[pdftex]{graphicx}
\usepackage{amsmath,amsfonts,amsbsy,amssymb,amsthm}
\usepackage{epstopdf}
\usepackage[caption=false]{subfig}
\usepackage{ulem,color}
\usepackage{enumitem}
\usepackage{algorithm}
\usepackage{algpseudocode}
\usepackage{siunitx}
\algrenewcommand{\algorithmicrequire}{\textbf{Input:}}
\algrenewcommand{\algorithmicensure}{\textbf{Output:}}

\newcommand{\ket}[1]{\left| #1 \right\rangle}
\newcommand{\bra}[1]{\left\langle #1 \right|}
\newcommand{\tr}[1]{\text{Tr}\left\{ #1 \right\}}

\begin{document}

\title{Maximizing device-independent randomness from a Bell experiment by optimizing the
  measurement settings}

\author{S~M~Assad}\email{cqtsma@gmail.com}
\author{O~Thearle}
\author{P~K~Lam}
\affiliation{Centre for Quantum Computation and Communication
  Technology, Department of Quantum Science, Research School of
  Physics and Engineering, The Australian National University,
  Canberra ACT 2601, Australia}

\begin{abstract}
  The rates at which a user can generate device-independent quantum
  random numbers from a Bell-type experiment depend on the
  measurements that he performs. By numerically optimising over these
  measurements, we present lower bounds on the randomness generation
  rates for a family of two-qubit states composed from a mixture of
  partially entangled states and the completely mixed state. We also
  report on the randomness generation rates from a tomographic
  measurement. Interestingly in this case, the randomness generation
  rates are not monotonic functions of entanglement.
\end{abstract}

\maketitle

\section{Introduction}
\label{intro}
Quantum mechanics is a probabilistic theory. It does not assign
definite outcomes to certain measurements. A physicist performing identical
measurements on two identically prepared systems might get different
measurement outcomes. Quantum mechanics postulates that the outcomes
of some measurements are undetermined before the measurement. This
randomness in the measurement outcomes has been used to generate
random numbers.

It might be argued that the randomness in the measurement outcome is
not really undetermined before the measurement. It is perhaps
determined by some hidden variables that provide a more complete
description of the system, but they are unknown to the
physicist. However, this hidden-variable description of nature was
recently tested in three Bell test experiments and was found to be
incompatible with the observed experimental
data~\cite{Hensen2015,Shalm2015,Giustina2015}. The observed data were
consistent with quantum mechanics. In other words, we see in our
experiments that nature behaves randomly, as postulated by quantum
mechanics. This implies that if the experimental observations obey
some relations and on the condition that the experiment was
performed correctly, we can certify the measurement outcomes were
undetermined before the measurement was performed. That is, their
outcomes generated new random numbers.

The conditions that need to be satisfied are those for a loophole-free
Bell experiment. Remarkably, these conditions do not include that the
physicist know the mechanisms of the measuring device. This
observation makes the realization of a device-independent (DI) quantum
random-number generator (QRNG) possible. In a DIQRNG, the user is able
to certify the creation of new random numbers despite being ignorant
of the device mechanisms.

In certifying the generation of new random numbers, the user trusts
that quantum mechanics provides a complete description of
nature. Based on the statistics of the measurement outcomes, he can
put a bound on the correlations between his measurement
outcomes and any other system that exists outside of his
lab\footnote{We assume that the laboratory is secure.}. This
bound allows him to extract new random numbers from the measurement
outcomes, that is, random numbers which are not correlated to any system
outside of his lab.

The first proof-of-concept DIQRNG used entangled photons generated in
an atomic ion trap to certify 42 new random numbers over a period of
about one month~\cite{Pironio2010}. More recently, using a more
efficient entanglement source, $4350$ bits of new randomness were
created at a rate of $0.4$ bits/s~\cite{Christensen2013}. Both setups
used the Clauser-Horne-Shimony-Holt (CHSH) value~\cite{Clauser1969} to
certify the randomness. The CHSH value is a function of the
measurement statistics, and this value sets a lower bound on the DI
randomness that can be certified. It turns out that using different
Bell operators, that is, different functions of the measurement
statistics, will give different equally valid lower bounds to the DI
randomness from the same measurement
statistics. In~\cite{Mironowicz2013}, several previously known as well
as \num{25000} randomly generated Bell operators were tested and shown to
certify varying amount of randomness from the two-qubit Werner
state. These operators were chosen in an {\it ad hoc} manner, and no single
operator was found to be optimal for all the Werner states.

In~\cite{Nieto-Silleras2014,Bancal2014}, the complete measurement
statistics were used to obtain a bound on the DI randomness instead of
resorting to a specific Bell operator. This gives the highest lower
bound on the DI randomness. A by-product of this process is
the optimal Bell operator that would have given the same
bound. This Bell operator gives the maximum DI randomness for the given
measurement statistics.

In a Bell setup for generating new random numbers, the physicist has a
choice of the measurement operators to use. By optimizing these
operators, he can get a better bound on the DI randomness. This is the
question that we address: How much randomness can the physicist
certify by using the optimal measurement operator? Recently, this question was
also addressed in~\cite{Mattar2015} for an experimentally
relevant optical Bell experiment setup and in~\cite{Passaro2015}, where
the requirement for full device independence was relaxed.

\section{Background}
We consider the usual Bell setup for generating DI random numbers. The
user inputs two random and independent measurement settings,
$x\in \left\{1,\ldots,M_x \right\}$ and $y\in \left\{1,\ldots,M_y
\right\}$, and receives two measurement
outcomes, $a,b \in \left\{-1,1\right\}$. In a DI setup, the user does
not have any knowledge of the measurement device. The behavior of the
apparatus is solely characterized by the conditional probabilities
$p(a,b|x,y)$, which we view as the components of the vector
$\mathbf{p}$. The user will use one measurement setting,
$(x^\ast, y^\ast)$, to generate his random numbers; the other
settings are only used to obtain bounds on the DI randomness.

Following~\cite{Nieto-Silleras2014,Bancal2014}, the maximum guessing probability for
an adversary, Eve, who is constrained by quantum mechanics and has
perfect knowledge of the measurement apparatus is
\begin{align}
&G\left[\mathbf{p} \right] = \max_{\{q_{ab},\mathbf{p}_{ab}\}} \sum_{ab}
q_{ab}  p_{ab}(a,b|x^\ast, y^\ast)\label{pri}\\
&\textrm{such that \;}  \sum_{ab} q_{ab}
  \mathbf{p}_{ab}=\mathbf{p}\;\label{con}\\
&\textrm{and \;} \mathbf{p}_{ab} \in \mathcal{Q}\;.\label{conQ}
\end{align}
The notation $\mathbf{p}\in \mathcal{Q}$ means that the conditional
probabilities $\mathbf{p}$ can be realized in quantum mechanics. In
other words, there exist a state $\rho$ and some measurement operators
$\pi_x^a$ and $\pi_y^b$ such that
$p(a,b|x,y)= \tr{\rho\, \pi_x^a \otimes \pi_y^b}$. The
constraint~(\ref{con}) ensures that the weighted sum of the particular
behaviors $\mathbf{p}_{ab}$ gives the observed behavior
$\mathbf{p}$. Eve can realize the guessing probability $G[\mathbf{p}]$
if the measurement device behaves according to $\mathbf{p}_{ab}$ with
probability $q_{ab}$ and Eve knows the particular behavior of each
measurement. For each instant of a particular behavior
$\mathbf{p}_{ab}$, Eve's guess of the measurement outcome will be
$(a,b)$. If the maximum guessing probability is less than 1, then
the lower bound to the amount of certifiable DI randomness is
quantified by the minimum entropy $H_\text{min}=-\log_2 G$.

The optimization problem~(\ref{pri}) is a conic linear program, and its
dual can be formulated as
\begin{align}
&D\left[\mathbf{p}\right] = \min_{\mathbf{f}} \mathbf{f}\cdot \mathbf{p} \label{dual}\\
&\textrm{such that \;} p'(a,b|x^\ast,y^\ast) \leq  \mathbf{f}\cdot
  \mathbf{p'} \;\textrm{ for }a,b \in  \left\{-1,1 \right\}\nonumber\\
&\textrm{and all \;} \mathbf{p'} \in \mathcal{Q}\;.\label{dualcon}
\end{align}
The solution of the dual program coincides with the solution of the
primal program: $D[\mathbf{p}]=G[\mathbf{p}]$. The optimization
variable $\mathbf{f}$ corresponds to a Bell expression that gives rise
to a guessing probability of $\mathbf{f}\cdot \mathbf{p}$. The optimal
$\mathbf{f}$ that achieves the minimum then corresponds to the optimal
Bell expression that minimizes Eve's guessing probability given the
behavior $\mathbf{p}$.

In general, the optimization problems (\ref{pri}) and (\ref{dual}) can
be computationally hard to solve. However the constraints (\ref{con})
and (\ref{dualcon}) can be relaxed \cite{Navascues2007,Navascues2008}
to give upper bounds to the guessing probabilities in a way that the
programs can be cast as a semidefinite program (SDP) which can be
solved efficiently. These relaxations can be progressively tightened
to give bounds that are successively tighter.

\section{Randomness maximization}
While the user of a DIRNG has no access to the workings of the device,
the physicist who builds the device has a choice of the quantum state
$\rho$ and the measurement operators $\boldsymbol{\pi}$ that he wants to
implement in the operation of the device. The vector
$\boldsymbol{\pi}$ has components
$\pi(a,b,x,y)=\pi_x^a\otimes\pi_y^b$ which are rank-one projectors
and satisfy
\begin{align}
\label{picon}
\begin{split}
&\tr{\pi_x^a\,
  \pi_x^{a'}}=\delta_{aa'} \;\textrm{ for }\;x \in
  \{1,\ldots,M_x\}\;\textrm{ and }\\
&\tr{\pi_y^b\,
  \pi_y^{b'}}=\delta_{bb'} \;\textrm{ for }\;y \in
  \{1,\ldots,M_y\}\;.
\end{split}
\end{align}
For example, if his machine can prepare the pure entangled two-qubit
state $\ket\psi=(\ket{00}+\ket{11})/\sqrt{2}$, then as shown
in~\cite{Dhara2013}, by designing the
measurement operators to be projectors along $\left(\ket{0}\cos
\alpha_x^a/2+\ket{1}\sin \alpha_x^a/2 \right)\otimes\left(\ket{0}\cos
\beta_y^b/2+\ket{1}\sin \beta_y^b/2\right)$ with the angles
\begin{equation}
\begin{aligned}
\label{m2}
\alpha_1^{ \pm1}&=\left(0,\pi\right)\,, & \alpha_2^{
  \pm1}&=\left(\frac{\pi}{2},-\frac{\pi}{2}\right)\,,
\\
\beta_1^{ \pm1}&=\left(\frac{\pi}{4},-\frac{3\pi}{4}\right)\,, &
\beta_2^{ \pm1}&=\left(\frac{3\pi}{4},-\frac{\pi}{4}\right)\,, &
\beta_3^{\pm1}&=\left(0,\pi\right)\, ,
\end{aligned}
\end{equation}
the device will be able to certify two bits of randomness with the
measurement settings $(2,3)$. However, if the measurement
operators used were not optimal, the machine will exhibit a different
behavior and may certify less randomness.

So if the builder can prepare a maximally entangled two-qubit state
and use the optimal measurement operator, then the device will be able
to certify two bits of randomness, and all is good. However, if the
builder is technologically limited to preparing some other state
$\rho$, then in general the measurement operators in~(\ref{m2}) will
not be optimal anymore. In this case, the builder is then interested
in finding the measurement operator he should implement that would
certify the maximum randomness\footnote{If some outcome symbols
  $(a,b)$ given $(x^*,y^*)$ are more unpredictable than others, then
  more randomness can potentially be extracted by post-selecting a
  subset of the symbols $(a,b)$. Although the post-selection reduce
  the number of data-points available for randomness extraction, the
  post-selected data might be more random, which makes it harder to
  for Eve to guess correctly. The net result can be an increase in the
  final randomness generation rate~\cite{Thinh2016}.} given that he is
restricted to the state $\rho$. This is the task that we shall now
investigate. More precisely, we want to find
\begin{align}
H\left[ \rho \right]=\max_{\boldsymbol{\pi}} D\left[p(\boldsymbol{\pi}) \right]\;,
\end{align}
where $p(\boldsymbol{\pi})=\tr{\rho \, \boldsymbol{\pi}}$ and the
vector $\boldsymbol{\pi}$ is constrained by~(\ref{picon}). Admittedly,
we have not solved this problem. Instead, we present and implement an
iterative algorithm in Algorithm~\ref{algo1} that converges to a
local maximum of $D\left[p(\boldsymbol{\pi}) \right]$.

\begin{algorithm}[H]
\caption{Proposed algorithm.}
\begin{algorithmic}[1]
 \Require input states $\rho$, initial positive operator-valued measure (POVM) $\boldsymbol{\pi}_0$, and
 stopping criteria $\epsilon$
 \State Initialize guessing probability $g_1=1$ and POVM $\boldsymbol{\pi}_1=\boldsymbol{\pi}_0$
 \Repeat
 \State Update $\boldsymbol{\pi}_0=\boldsymbol{\pi}_1$ and $g_0=g_1$
 \State Compute $\mathbf{p}=p(\boldsymbol{\pi}_0)$
 \State Compute $D[\mathbf{p}]$ and corresponding $\mathbf{f}$ by solving the relaxed version of~(\ref{dual})
 \State Compute  the minimum of $g_1(\boldsymbol{\pi})=\mathbf{f}\cdot
 p\left(\boldsymbol{\pi}_1\right)$ and corresponding $\boldsymbol{\pi}_1$\label{step5}
 \Until{$g_0-g_1\le \epsilon$}
 \Ensure $\boldsymbol{\pi}_1$
\end{algorithmic}
\label{algo1}
\end{algorithm}
The tolerance $\epsilon$ sets the stopping condition for the
algorithm. In step~\ref{step5}, we compute the minimum of guessing
probability $\mathbf{f}\cdot p\left(\boldsymbol{\pi}\right)$ which
corresponds to finding the measurement settings that maximizes the
Bell value for a given Bell expression $\mathbf{f}$. The guessing
probability
$\mathbf{f}\cdot
p\left(\boldsymbol{\pi}\right)=\tr{\rho\,\mathbf{f}\cdot
  \boldsymbol{\pi} }$ is a quadratic function of $\pi_x^a$ and
$\pi_y^b$ with the quadratic constraints~(\ref{picon}). We can use the
Lagrange multiplier method to find the minimum.

While the algorithm might not find the global maximum
$H[\rho]$, it usually finds measurement settings that yield more DI
randomness than a randomly chosen measurement setting. In our
implementation, we use several initial settings $\boldsymbol{\pi}_0$ in an
attempt to find the global maximum. All SDP calculations were
performed using the CVX package for MATLAB~\cite{Grant2008,Grant2014}.

\begin{figure}
\newlength\figHeight 
\newlength\figWidth 
\includegraphics[width=1.\linewidth]{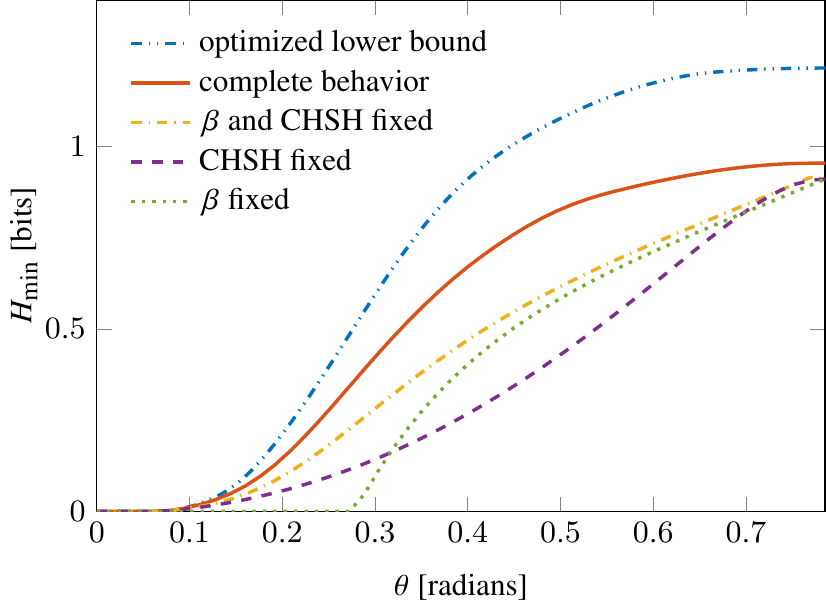}
\caption{{\bf Comparison of lower bounds on DI randomness for
    $v=0.99$.} The green dotted and purple dashed lines show the DI
  randomness obtained when constrained by the Bell
  operators~(\ref{Ibeta}) and~(\ref{ICHSH}) with a fixed measurement
  direction~\cite{Nieto-Silleras2014}. Using both operators together
  gives a higher randomness, depicted by the yellow dash-dotted
  line. Constraining Eve to the complete behavior gives the most
  randomness from the fixed behavior generated from the measurement
  direction depicted by the solid orange
  line~\cite{Nieto-Silleras2014}. The top line denotes the randomness
  bound for an optimized measurement direction. These curves were
  obtained with a third-order relaxation of the SDP
  hierarchy~\cite{Navascues2007,Navascues2008}.}
\label{fig:hmin99}
\end{figure}

\section{Results}
We apply our algorithm to the family of states
\label{rho}
\begin{align}
\rho(v,\theta)= v \ket{\Psi_\theta}\bra{\Psi_\theta} + (1-v)\frac{1}{4}\;,
\end{align}
where $\ket{\Psi_\theta}=\ket{00}\cos\theta+\ket{11}\sin\theta$ and
visibility $0\leq v \leq 1$ gives the fraction of the state
$\ket{\Psi_\theta}$. In the noiseless limit of $v=1$,
arbitrarily close to two bits of DI randomness can be attained in the
maximally entangled case when $\theta=\pi/4$ with $M_x=M_y=2$ measurement
settings~\cite{Acin2012}. Two bits of DI randomness are
also achievable when $\theta$ is arbitrarily close to zero with $M_x=M_y=4$
measurement settings~\cite{Acin2012}.

We first consider the case where $M_x=M_y=2$ and the visibility is
fixed at $v=0.99$. In Fig.~\ref{fig:hmin99}, we compare the
DI randomness from the optimized measurement setting to a bound obtained
using a fixed measurement setting as reported
in~\cite{Nieto-Silleras2014}. We see a significant improvement in the
certifiable randomness using the optimized measurement settings. For
completeness, we also include the certifiable randomness constrained using
two specific Bell operators
\begin{align}
 I_\textrm{CHSH}&=\left< A_1 B_1 \right>+\left< A_1 B_2 \right>+\left<
  A_2 B_1 \right>-\left< A_2 B_2 \right>\;\textrm{ and }\label{ICHSH}\\
I^\beta &= I_\textrm{CHSH} + \beta \left< A_1 \right>\;\label{Ibeta}
\end{align}
and also constrained by both operators
together~\cite{Nieto-Silleras2014} using a fixed measurement setting where
$\beta=2\cos 2\theta/\sqrt{1+\sin^2 2\theta}$, $\left<A_x
  B_y\right>=\sum_{ab}a b \,p(a,b|x,y)$, and $\left<A_x \right>=\sum_a
a\, p(a|x)$. The DI randomness bounds using specific operators are always
lower than using the complete measurement statistics.

\begin{figure}
\includegraphics[width=1.\linewidth]{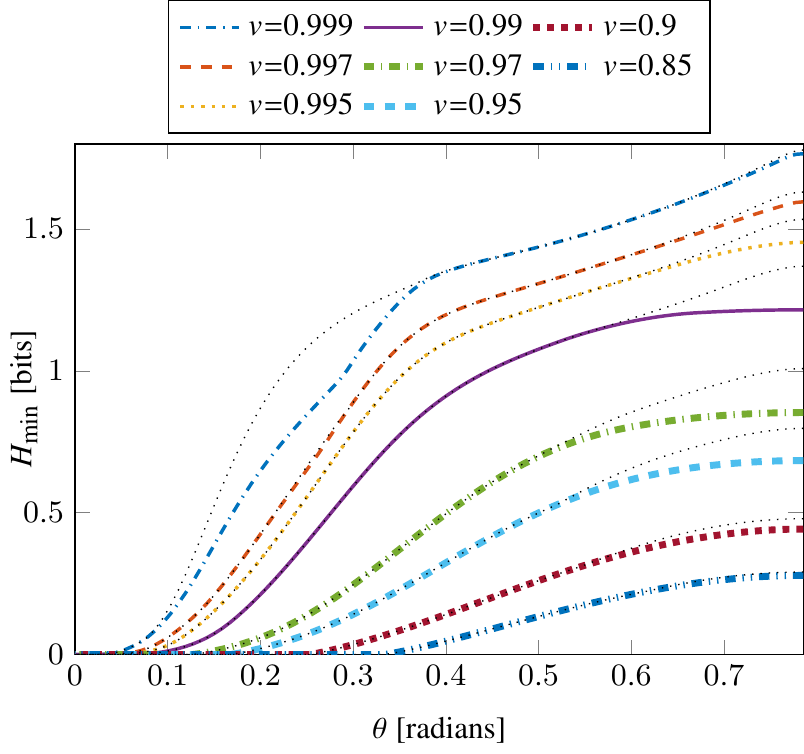}
\caption{{\bf Optimized lower bounds on DI randomness for various
    visibilities with two measurement settings for each side.} The
  black dots have four measurement settings for each side. These
  curves were obtained with a second-order relaxation of the SDP
  hierarchy.}
\label{fig:hmin22}
\end{figure}

\begin{figure}
\includegraphics[width=1.\linewidth]{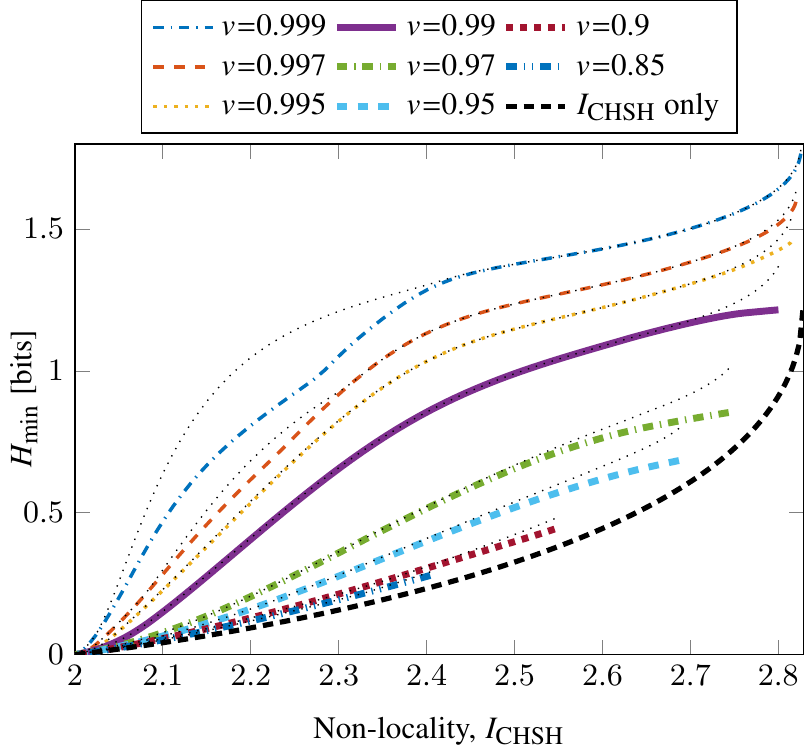}
\caption{{\bf Optimized lower bounds on DI randomness as a function of
    nonlocality.} Different amount of randomness from the same amount
  of CHSH violation with two measurement settings for each
  side. Black dots have four measurement settings for each
  side. The lowest dashed line shows the DI randomness bound obtained
  from using the CHSH value alone. These curves were obtained with a
  second-order relaxation of the SDP hierarchy.}
\label{fig:chsh}
\end{figure}

Next, we plot the DI randomness bound as a function of $\theta$ for
various visibilities in Fig.~\ref{fig:hmin22} for $M_x=M_y=2$. We
also plotted the DI randomness when $M_x=M_y=4$ in the same figure. In
most cases, the improvement obtained from using four measurement
settings is not very significant. In Fig.~\ref{fig:chsh}, we plot
the DI randomness as a function of nonlocality as
measured by the CHSH value $I_\textrm{CHSH}$. Relying on the CHSH value alone
gives a much lower DI randomness, especially when the state has a high
visibility. Even with a maximally entangled two-qubit state, a CHSH
value of $2\sqrt{2}$ can only certify $1.22845$ bits of randomness.

\begin{figure}
\includegraphics[width=1.\linewidth]{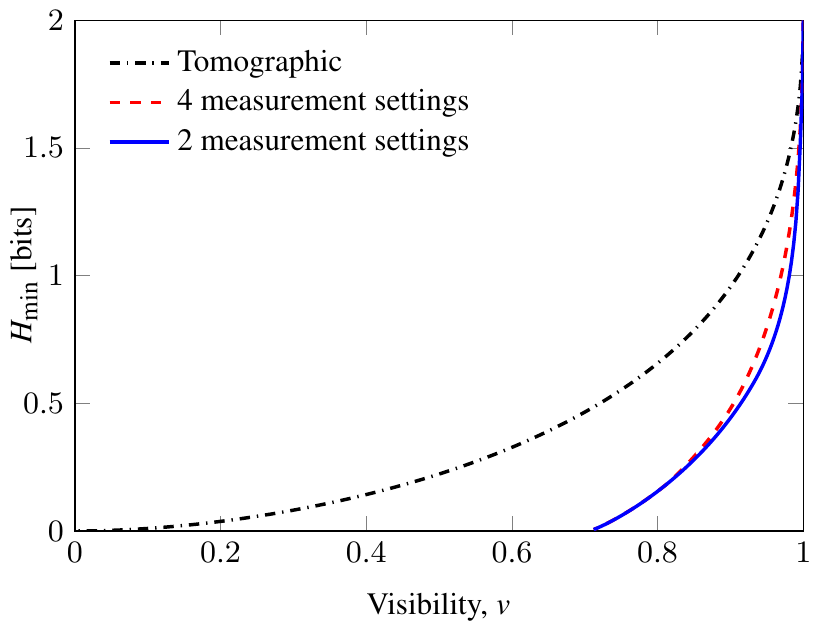}
\caption{{\bf Optimized lower bounds on DI randomness as a function of
    visibility from a mixture of a maximally entangled state and white
    noise.} The DI randomness goes to zero when $v<1/\sqrt{2}$ for two
  (solid line) and four (dashed line) measurement settings on each
  side.  The four-measurement-setting randomness bound that we report
  here is slightly higher than the results reported in~\cite{Law2014},
  where there are two fixed settings for one side and four fixed
  settings for the other side. We computed the fixed settings using
  both the second- and third-order relaxations of the SDP hierarchy,
  but they might turn out to be identical when a tighter constraint is
  used. We find no improvement in the tomographic result (dash-dotted
  line) compared to the results using a fixed measurement setting
  reported in~\cite{Law2014}. The two-setting and four-setting curves
  were obtained using a second-order relaxation of the SDP hierarchy.}
\label{fig:vis}
\end{figure}

In Fig.~\ref{fig:vis}, we fix the input state to have $\theta=\pi/4$
and plot the DI randomness as a function of visibility for $M_x=M_y=2$
and $M_x=M_y=4$. There is only a slight increase in the DI
randomness bound when going to four measurement settings. The DI
randomness increases monotonically with $v$ as one would expect. This
is because from a high-visibility state, one can always introduce noise
to get to a state with lower visibility and attain at least the same DI randomness.

\begin{figure}
\includegraphics[width=1.\linewidth]{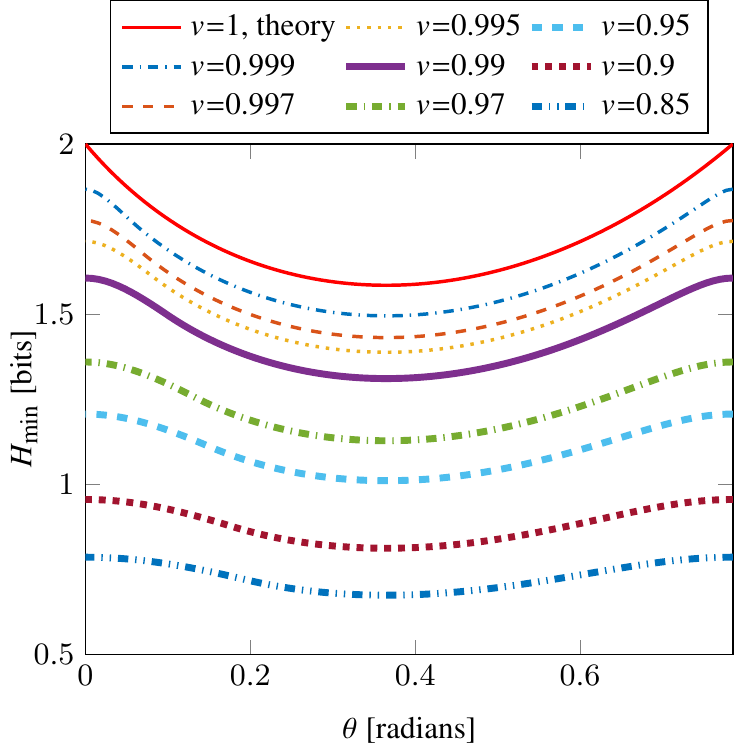}
\caption{{\bf Optimized randomness with complete tomography.} With complete
  tomography, the randomness generation rate is not zero even when the
  two-qubit state is separable at $\theta=0$. For a fixed visibility,
  the randomness rate is not a monotonic function of $\theta$. It is
  maximum when $\theta=0$ and $\theta=\pi/4$.}
\label{fig:tomo}
\end{figure}

\begin{figure}
\setlength\figHeight{.8\linewidth} 
\setlength\figWidth{.8\linewidth}
\centering
\includegraphics[width=1.\linewidth]{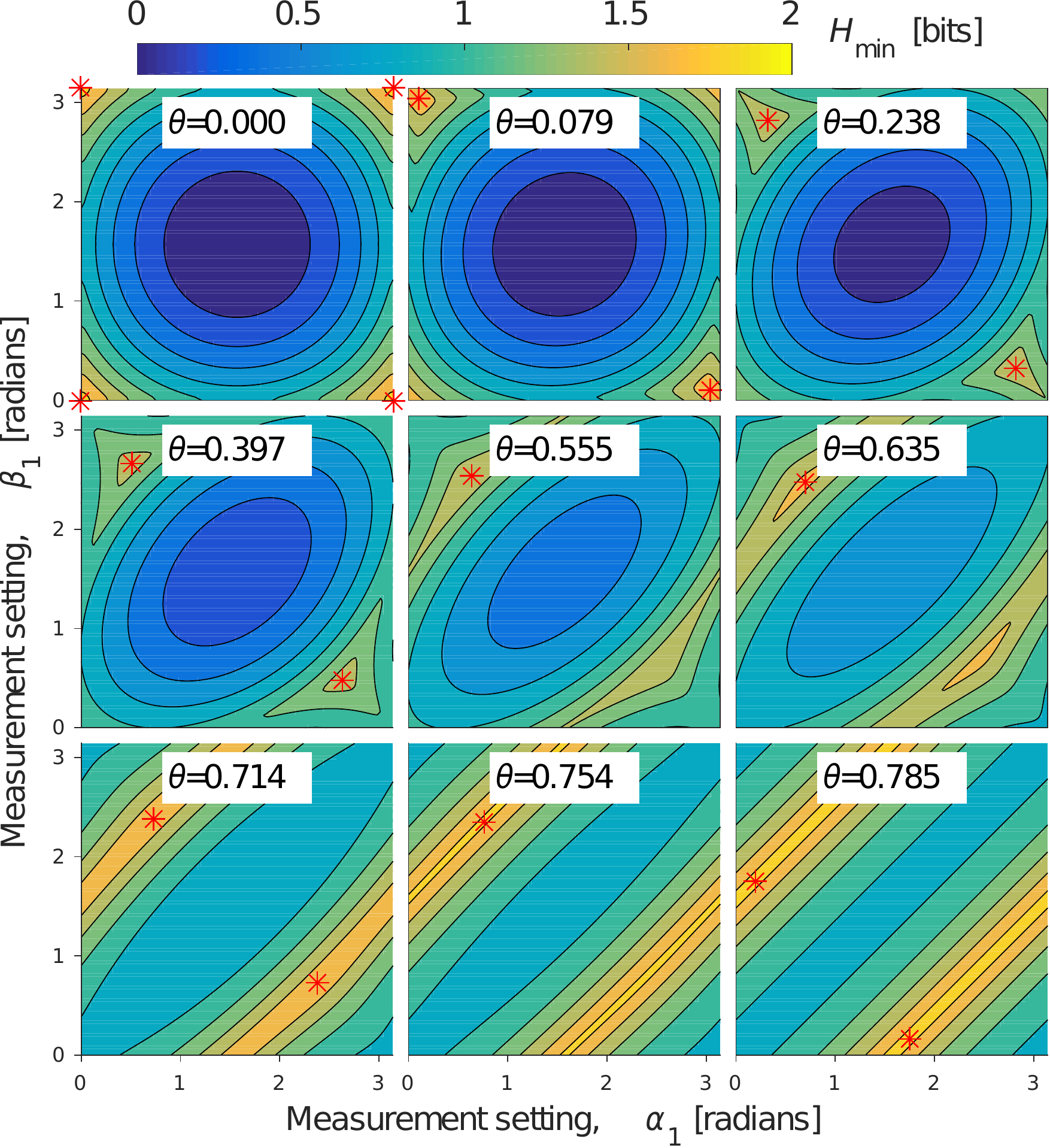}
\caption{{\bf Randomness with complete tomography as a function of
    measurement angles.} $H_\textrm{min}$ as a function of the
  measurement settings for different input states parametrized by
  $\theta$ from $0$ to $\pi/4$ with $v=0.999$. The $x$ axis corresponds
  to the angle $\alpha_1$ of the Bloch vector
  $(\sin\alpha_1, 0, \cos\alpha_1)$ of the measurement setting for the
  first side, and the $y$ axis corresponds to the angle $\beta_1$ of the
  Bloch vector $(\sin\beta_1, 0, \cos\beta_1)$ of the measurement setting for
  the second side. The red asterisk denotes the maximum
  $H_\textrm{min}$ value for each $\theta$. }
\label{fig:tomo_map}
\end{figure}

Finally, in the limit when the number of settings becomes large, the
DI randomness will be upper bounded by the setup where
the user can perform a complete tomography~\cite{Law2014}. In
this case, the constraint~(\ref{con}) is replaced by a constraint on
the quantum states $\rho_{ab}$ with
$\sum_{ab}q_{ab}\rho_{ab}=\rho$. The constraints that $\rho_{ab}$ is
positive mean that programs~(\ref{pri}) and~(\ref{dual}) are already
SDPs.  We plot the tomographic randomness rate in
Fig.~\ref{fig:tomo}. For a fixed
$\theta$, the tomographic randomness rates decrease with
$v$. However, the tomographic randomness rates
are not monotonic in $\theta$ for a fixed $v$. For the same $v$, starting with a
state with small entanglement (low $\theta$) can still yield
the same amount of randomness as a state with large entanglement
($\theta$ near $\pi/4$). The dip in the randomness rates when
$\theta=\pi/8$ is unlikely due to the algorithm
being stuck in a local maximum. We check this numerically by scanning
the whole parameter space. For the case
of a qubit pair input that we are considering, the measurement directions
that the user uses to generate his tomographic randomness can be
parametrized by the Bloch vector angles $\alpha_1$ and $\beta_1$. 
Some typical tomographic randomness rates are shown in Fig.~\ref{fig:tomo_map}
as a function of the two Bloch vectors.

In Fig.~\ref{fig:vis}, we plot the randomness from a tomographic
measurement when $\theta=\pi/4$ as a function of $v$.  We find no
improvement compared to the results reported
in~\cite{Law2014}. The measurement used there,
\begin{align}
\alpha_1^{ \pm1}&=\left(0,\pi\right)\textrm{\; and \;} & \beta_1^{ \pm1}&=\left(\frac{\pi}{2},-\frac{\pi}{2}\right)\;,
\end{align}
indeed attains the maximum randomness we found. 

When the visibility is exactly unity, the quantum state that the user has
is a pure state. For this, Eve's guessing probability can be
calculated exactly and then maximized over the user's
measurements. The final result is
\begin{align}
G = \frac{1}{4}\left(1+\sin 2\theta \right)\cos^2 \alpha \;,
\end{align}
where $\alpha$ characterizes the measurement direction and is given by solving
\begin{align}
\sin \alpha =
  \frac{-\cos 2\theta+\sqrt{\cos^2 2\theta +4\sin 2\theta (1+\sin
  2\theta)}}{2\left(1+\sin 2\theta\right)}\;.
\end{align}
The min-entropy from this guessing probability is plotted as the top
line in Fig.~\ref{fig:tomo}. We see that two bits of randomness are
achievable only when the state is maximally entangled or when it is
separable.

\section{Conclusions}
The amount of randomness generated from a DIQRNG can be improved by
optimizing the measurement setting. However, for the two-qubit state
considered, the additional improvement achieved by using four
measurement settings on each side is in most cases not
significant. There is a disadvantage in having more measurement settings:
the experimental setup is more complicated and more data are needed to
characterize the measurements. This is not justified by the minimal
increase in the randomness generation rates.

\begin{acknowledgments}
We acknowledge the support of the Australian Research
Council Centre of Excellence for Quantum Computation and Communication
Technology (Project No. CE110001027).
\end{acknowledgments}


\begin{thebibliography}{100}

\bibitem{Hensen2015}
B.~Hensen, H.~Bernien, A.~E. Dr\'{e}au, A.~Reiserer, N.~Kalb, M.~S. Blok,
  J.~Ruitenberg, R.~F.~L. Vermeulen, R.~N. Schouten, C.~Abell\'{a}n, W.~Amaya,
  V.~Pruneri, M.~W. Mitchell, M.~Markham, D.~J. Twitchen, D.~Elkouss,
  S.~Wehner, T.~H. Taminiau, and R.~Hanson.
\newblock {Loophole-free Bell inequality violation using electron spins
  separated by 1.3 kilometres}.
\newblock {\em Nature}, 526(7575):682--686, October 2015.

\bibitem{Shalm2015}
Lynden~K. Shalm, Evan Meyer-Scott, Bradley~G. Christensen, Peter Bierhorst,
  Michael~A. Wayne, Martin~J. Stevens, Thomas Gerrits, Scott Glancy, Deny~R.
  Hamel, Michael~S. Allman, Kevin~J. Coakley, Shellee~D. Dyer, Carson Hodge,
  Adriana~E. Lita, Varun~B. Verma, Camilla Lambrocco, Edward Tortorici, Alan~L.
  Migdall, Yanbao Zhang, Daniel~R. Kumor, William~H. Farr, Francesco Marsili,
  Matthew~D. Shaw, Jeffrey~A. Stern, Carlos Abell\'an, Waldimar Amaya, Valerio
  Pruneri, Thomas Jennewein, Morgan~W. Mitchell, Paul~G. Kwiat, Joshua~C.
  Bienfang, Richard~P. Mirin, Emanuel Knill, and Sae~Woo Nam.
\newblock Strong loophole-free test of local realism.
\newblock {\em Phys. Rev. Lett.}, 115:250402, Dec 2015.

\bibitem{Giustina2015}
Marissa Giustina, Marijn A.~M. Versteegh, S\"oren Wengerowsky, Johannes
  Handsteiner, Armin Hochrainer, Kevin Phelan, Fabian Steinlechner, Johannes
  Kofler, Jan-\AA{}ke Larsson, Carlos Abell\'an, Waldimar Amaya, Valerio
  Pruneri, Morgan~W. Mitchell, J\"orn Beyer, Thomas Gerrits, Adriana~E. Lita,
  Lynden~K. Shalm, Sae~Woo Nam, Thomas Scheidl, Rupert Ursin, Bernhard
  Wittmann, and Anton Zeilinger.
\newblock Significant-loophole-free test of Bell's theorem with entangled
  photons.
\newblock {\em Phys. Rev. Lett.}, 115:250401, Dec 2015.

\bibitem{Pironio2010}
S.~Pironio, A.~Acin, S.~Massar, A.~Boyer de~la Giroday, D.~N. Matsukevich,
  P.~Maunz, S.~Olmschenk, D.~Hayes, L.~Luo, T.~A. Manning, and et~al.
\newblock Random numbers certified by Bell's theorem.
\newblock {\em Nature}, 464(7291):1021--1024, Apr 2010.

\bibitem{Christensen2013}
B.~G. Christensen, K.~T. McCusker, J.~B. Altepeter, B.~Calkins, T.~Gerrits,
  A.~E. Lita, A.~Miller, L.~K. Shalm, Y.~Zhang, S.~W. Nam, and et~al.
\newblock Detection-loophole-free test of quantum nonlocality, and
  applications.
\newblock {\em Phys. Rev. Lett.}, 111(13), Sep 2013.

\bibitem{Clauser1969}
John~F. Clauser, Michael~A. Horne, Abner Shimony, and Richard~A. Holt.
\newblock Proposed experiment to test local hidden-variable theories.
\newblock {\em Phys. Rev. Lett.}, 23(15):880--884, Oct 1969.

\bibitem{Mironowicz2013}
Piotr Mironowicz and Marcin Pawlowski.
\newblock Robustness of quantum-randomness expansion protocols in the presence
  of noise.
\newblock {\em Phys. Rev. A}, 88(3), Sep 2013.

\bibitem{Nieto-Silleras2014}
O~Nieto-Silleras, S~Pironio, and J~Silman.
\newblock Using complete measurement statistics for optimal device-independent
  randomness evaluation.
\newblock {\em New J. Phys.}, 16(1):013035, Jan 2014.

\bibitem{Bancal2014}
Jean-Daniel Bancal, Lana Sheridan, and Valerio Scarani.
\newblock More randomness from the same data.
\newblock {\em New J. Phys.}, 16(3):033011, Mar 2014.

\bibitem{Mattar2015}
Alejandro Mattar, Paul Skrzypczyk, Jonatan~Bohr Brask, Daniel Cavalcanti, and
  Antonio Acin.
\newblock Optimal randomness generation from optical Bell experiments.
\newblock {\em New J. Phys.}, 17(2):022003, Feb 2015.

\bibitem{Passaro2015}
Elsa Passaro, Daniel Cavalcanti, Paul Skrzypczyk, and Antonio Ac\'{i}n.
\newblock Optimal randomness certification in the quantum steering and
  prepare-and-measure scenarios.
\newblock {\em New J. Phys.}, 17(11):113010, Oct 2015.

\bibitem{Navascues2007}
Miguel Navascu\'es, Stefano Pironio, and Antonio Ac\'{i}n.
\newblock Bounding the set of quantum correlations.
\newblock {\em Phys. Rev. Lett.}, 98:010401, Jan 2007.

\bibitem{Navascues2008}
Miguel Navascu\'es, Stefano Pironio, and Antonio Ac\'in.
\newblock A convergent hierarchy of semidefinite programs characterizing the
  set of quantum correlations.
\newblock {\em New J. of Phys.}, 10(7):073013, Jul 2008.

\bibitem{Dhara2013}
Chirag Dhara, Giuseppe Prettico, and Antonio Ac\'{i}n.
\newblock Maximal quantum randomness in Bell tests.
\newblock {\em Phys. Rev. A}, 88:052116, Nov 2013.

\bibitem{Grant2008}
Michael Grant and Stephen Boyd.
\newblock Graph implementations for nonsmooth convex programs.
\newblock In V.~Blondel, S.~Boyd, and H.~Kimura, editors, {\em Recent Advances
  in Learning and Control}, Lecture Notes in Control and Information Sciences,
  pages 95--110. Springer-Verlag Limited, 2008.
\newblock \url{http://stanford.edu/~boyd/graph_dcp.html}.

\bibitem{Grant2014}
Michael Grant and Stephen Boyd.
\newblock {CVX}: Matlab software for disciplined convex programming, version
  2.1.
\newblock \url{http://cvxr.com/cvx}, March 2014.

\bibitem{Acin2012}
Antonio Ac\'{i}n, Serge Massar, and Stefano Pironio.
\newblock Randomness versus nonlocality and entanglement.
\newblock {\em Phys. Rev. Lett.}, 108:100402, Mar 2012.

\bibitem{Law2014}
Yun~Zhi Law, Le~Phuc Thinh, Jean-Daniel Bancal, and Valerio Scarani.
\newblock Quantum randomness extraction for various levels of characterization
  of the devices.
\newblock {\em J. Phys. A: Math. Theor.}, 47(42):424028, Oct 2014.

\bibitem{Thinh2016}
Le~Phuc Thinh, Gonzalo de~la Torre, Jean-Daniel Bancal, Stefano Pironio, and
  Valerio Scarani.
\newblock Randomness in post-selected events.
\newblock {\em New J. Phys.}, 18(3):035007, Mar 2016.

\end{thebibliography}
\end{document}